\documentclass[]{jfm}

\usepackage{bm} 
\usepackage{amssymb}
\usepackage{natbib}
\usepackage{graphicx}
\usepackage{color}
\usepackage{amsmath}
\usepackage{amsfonts}
\usepackage{bm}

\newcommand{\br}{{\bf r}}  
  
\newcommand{\bx}{{\bf x}}  
\newcommand{\bX}{{\bf X}}  
\newcommand{\cS}{{\cal S}}  
\newcommand{\cF}{{\cal F}}

\newcommand{\bu}{{\bf u}}

\newcommand{\ee}{\end{equation}}
\newcommand{\be}{\begin{equation}}

\usepackage[latin1]{inputenc}
\newcommand{\beq}{\begin{equation}}
\newcommand{\eeq}{\end{equation}}
\newcommand{\bea}{\begin{eqnarray}}
\newcommand{\eea}{\end{eqnarray}}

\usepackage{amsmath}
\usepackage{amsfonts}
\usepackage{amssymb}
\usepackage{bm}
\usepackage{color}
\usepackage{graphicx}

\begin{document}
\title[Eulerian and Lagrangian Statistics in weakly compressible
turbulence]{Eulerian and Lagrangian Statistics from high resolution
  Numerical Simulations of weakly compressible turbulence}

\author[ R. Benzi, L. Biferale, R. Fisher, D.Q. Lamb and F. Toschi]
{R. BENZI$^{1}$, L. BIFERALE$^{1}$, R. FISHER$^2$, D. Q. LAMB$^3$ and\\
  F. TOSCHI$^{4}$}

\affiliation{$^1$Department of Physics and INFN, University of Rome Tor Vergata, \\ Via della Ricerca Scientifica 1, 00133 Rome, Italy\\
  $^2$ Department of Physics, University of Massachusetts at
  Dartmouth,\\ 285 Old Westport Road, Dartmouth, Ma. 02740\\
  $^3$ Center for Astrophysical Thermonuclear Flashes, The University
  of Chicago, Chicago, Illinois 60637, USA and Department of Astronomy
  and Astrophysics, The University of Chicago, Chicago, Illinois
  60637, USA\\
  $^4$ Department of Physics and Department of Mathematics and
  Computer Science, Eindhoven University of Technology, 5600 MB
  Eindhoven, The Netherlands and Istituto per le Applicazioni del
  Calcolo CNR, Viale del Policlinico 137, 00161 Roma, Italy}

\date{\today}
\maketitle

\begin{abstract}
  We report a detailed study of Eulerian and Lagrangian statistics
  from high resolution Direct Numerical Simulations of isotropic
  weakly compressible turbulence.  Reynolds number at the Taylor
  microscale is estimated to be around $600$. Eulerian and Lagrangian
  statistics is evaluated over a huge data set, made by $1856^3$
  spatial collocation points and by 16 million particles, followed
  for about one large-scale eddy turn over time. We present data for
  Eulerian and Lagrangian Structure functions up to ten order. We
  analyse the local scaling properties in the inertial range and in
  the viscous range. Eulerian results show a good superposition with previous data. Lagrangian statistics is different from existing  experimental and numerical results, for moments of sixth order and 
higher. 
We interpret this in terms of a
possible contamination from viscous scale affecting the  
estimate of the scaling properties in previous studies. 
We show that a simple bridge relation based on
  Multifractal theory is able to connect scaling properties of both
  Eulerian and Lagrangian observables, provided that the small
  differences between intermittency of transverse and longitudinal
  Eulerian structure functions are properly considered.
\end{abstract}

\section{Introduction}
In the last few years many interesting and remarkabe  results have been
obtained by investigating the statistical properties of Lagrangian
particles advected by a turbulent flow --for a recent review see
\cite{toschibod}. For Lagrangian particles we mean simple point like
ideal tracers, whose instantaneous velocity coincides with the local
Eulerian velocity field, $\bu(\bx,t)$: $\dot \bX(t) =
\bu(\bX(t),t)$. Beside being relevant in many applications, where
transport and/or aggregation of particles is important, the study of
fully developed turbulence in the Lagrangian framework has opened
important new directions of scientific investigations. In
particular, small scale intermittency of turbulent flows and
dissipation range can be probed more efficiently in the Lagrangian
framework as shown experimentally since the works of
\cite{LPVCAB.01,MMMP.01,mordant,ott_mann,berg,bourgoin,tsinober,luthi2} and numerically in
\cite{BBCLT.05,pkyeung,mueller,mordant}, (see also \cite{biferaletal07} and
\cite{arneodoetal08} for comparison between experimental and numerical
Lagrangian data). Moreover, Lagrangian and Eulerian measurement (those
on the reference frame of the laboratory) should be  intimately statistically
connected as pointed out for the first time in \cite{Bo.93,bof02}, the hope is to learn about the former
from the study of the latter and viceversa.

In this paper we present some new results concerning the statistical
properties of the Lagrangian and Eulerian velocity field in weakly
compressible turbulent flow. The flow is the outcome of a high
resolution numerical simulations described  in the next
section.  Our motivation is threefold: due to the high spatial resolution and
the huge particle statistics (16 millions of particles) we are able to extract the statistics of the
Lagrangian velocity field with high accuracy and relatively small
error bars. From this point of view we are able to assess, for the
first time at this Reynolds numbers, statistical fluctuations in the Lagrangian domain as
intense as those given by moments up to order 10. The second purpose
is to understand whether the effect of the weak compressibility
changes the statistical properties of Eulerian and  Lagrangian velocity field with
respect to previous numerical investigations of incompressible
isotropic turbulent flows.  Preliminary results concerning this point
have been already published in \cite{prl.chicago}; here a more
detailed analysis on new observables is reported. In particular, we
are interested in understanding the effect of intermittency in the
inertial range for Lagrangian particles. Some results concerning
dissipative scales will be also presented.  Finally, our third
motivation  concerns
with the statistical link between Lagrangian and Eulerian
measurements. Concerning this point we show that a simple model, based
on the multifractal theory, is able to translate between the two
ensembles with a quantitative agreement extending over three decades
from inside dissipative scales to the integral scale (\cite{arneodoetal08}).
 
Our main tool is the numerical computation of velocity increments
over a time lag $\tau$, the so-called Lagrangian structure functions:
\begin{equation}
  \cS_i^{(p)}(\tau) = \langle [v_i(t+\tau) -v_i(t)]^p \rangle=
  \langle (\delta_\tau v_i)^p \rangle, 
\label{eq:LSF}
\end{equation} 
where $i=x,y,z$ are the three velocity components along a particle
trajectory, $v_i(t) = u_i({\bx}(t),t)$ and the average is defined over
the ensemble of particle trajectories.  As stationary and homogeneity
is assumed, moments of velocity increments only depend on the time lag
$\tau$. When isotropy is also valid, all components must
be symmetric and we will drop the dependency on the spatial index. In
the inertial range, for time lags smaller than the integral
time and larger than the Kolmogorov time, $\tau_\eta \ll \tau \ll
T_L$, non-linear energy transfer governs the dynamics.  From a
dimensional viewpoint, only the scale $\tau$ and the energy transfer
$\epsilon$ should enter the structure functions. The only admissible
choice in isotropic statistics is $\cS^{(p)}(\tau) \sim (\epsilon
\tau)^{p/2}$, but it does not take into account the fluctuating nature
of dissipation.  Many empirical studies (\cite{toschibod}) have indeed
shown that the tail of the probability density functions of
$\delta_\tau v$ become increasingly non-Gaussian at decreasing
$\tau/T_L$, leading to intermittency and to anomalous scaling
exponents, meaning a breakdown of the dimensional law, i.e.
\begin{equation}
  \cS^{(p)}(\tau)  \sim \tau^{\xi^{(p)}}\,
  \label{eq:scaling}
\end{equation}
with $\xi^{(p)}\neq p/2$.

As we will review and test in details later, the Lagrangian
measurements can be linked to Eulerian Structure Function (ESF). In
isotropic statistics we may have two different Eulerian increments,
longitudinal and transverse (see \cite{Fr.95} for a textbook
introduction to scaling in homogeneous and isotropic turbulence): \be
\begin{cases}
S^{(p)}_L(r) \equiv \langle [\bu(\bx+\br)-\bu(\bx)] \cdot {\hat
  \br}]^p \rangle\\
 S^{(p)}_T(r) \equiv \langle 
|\bu(\bx+{\br}_T)-\bu(\bx)|^p \rangle,
\qquad {\br}_T \cdot {\bu} = 0.
\end{cases}
\ee It is well established, (\cite{arneodoetal96,Fr.95}) that also
Eulerian statistics have anomalous scaling, for $\eta \ll r \ll L$:
\be
\begin{cases}
  S^{(p)}_L(r) \sim r^{\zeta_L^{(p)}} \\
  S^{(p)}_T(r) \sim r^{\zeta_T^{(p)}}. \\
\end{cases}
\ee From a theoretical point of view, in isotropic turbulence one
would expect that longitudinal and transverse fluctuations have the
same scaling, $\zeta_L^{(p)} = \zeta_T^{(p)}$ (see \cite{bp} for a review on
anisotropic turbulence).  This is not what was observed in many
experimental and numerical analysis
(\cite{boratav,vander,shen-warhaft,chen,zhou,dhruva}). Discrepancy can be due to
finite Reynolds numbers (\cite{Hill,he}) or to some remnant
small-scale anisotropy (\cite{bp}). Still, even large Reynolds number
DNS with isotropic forcing show some persistent differences between
$\zeta_L^{(p)}$ and $\zeta_T^{(p)}$ as shown in \cite{toshi.arfm,Go.02}; the
same discrepancy is quantitatively confirmed in the analysis here
reported. Whether this effect will persist at higher Reynolds
number remains an important open question for further experimental and
numerical investigations.

Recently, it has been argued that there must exists a relation between
Lagrangian and Eulerian exponents, on the basis of a common
multifractal description (\cite{bof02,BBCLT.05}).  This relation
between Lagrangian and Eulerian statistical properties is highly non
trivial and it is worth to understand how accurately can be verified
in the existing DNS and laboratory experiments. It has been proved to
be very efficient to predict the pdf of Lagrangian acceleration
(\cite{BBCDLT.04}) and of Lagrangian Structure Functions of low order
(\cite{arneodoetal08}).

The paper is organized as follows. In section (\ref{sec:dns}) we
discuss the numerical simulations and the data set. In section
(\ref{sec:3}) we discuss the computation of the Eulerian structure
functions.  In section (\ref{sec:4}) we report the analysis of
Lagrangian structure functions and we discuss the physical results
obtained by our analysis. In section (\ref{sec:5}) we provide a
theoretical interpretation of our findings by employing the
multifractal framework. Conclusions follows in section (\ref{sec:6}).

\section{Data set and numerical simulations}
\label{sec:dns}
\begin{figure}
  \begin{center}
    \includegraphics[width=0.8\textwidth]{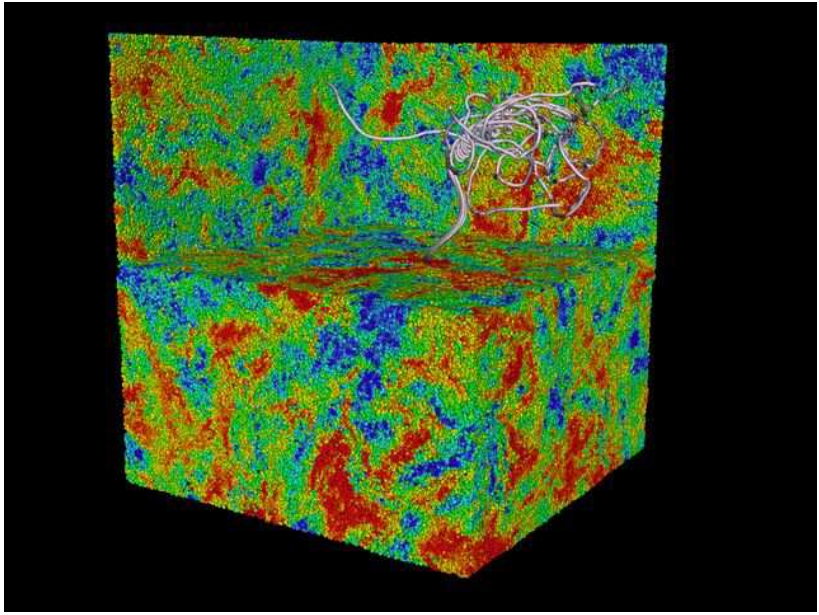}
    \caption{Example of Eulerian and Lagrangian rendering. Both the
      intensity of the Eulerian velocity field at a given time and the
      Lagrangian evolution of a bunch of particles with trajectories
      ending at the time of the Eulerian snapshot are shown. Notice
      that the Lagrangian particles have an initial smooth evolution
      because we also show the initial transient time when the
      underlying Eulerian field was chosen smooth and with low
      energy. The bunch of particles was chosen such has to encounter
      a vortex filaments during their evolution. Figure courtesy of
      B. Gallagher.}
    \label{fig.k45}
  \end{center}
\end{figure}

As briefly highlight in the introduction, our main purpose is to
investigate intermittency of the Lagrangian/Eulerian velocity field in
the inertial range for a weakly compressible isotropic turbulent flow
(see figure (\ref{fig.k45}) for a colorful rendering).  The numerical
simulation have been performed using the FLASH code developed by the
Flash Center at the University of Chicago (\cite{chicago}). The
equation of motion are:
\begin{equation}
  \label{eq:eq}
  \begin{cases}
    {\partial_t \rho} + {\bm \nabla} \cdot ( \bu \rho) = 0 \\
    {\partial_t (\rho \bu)} + {\bm \nabla} \cdot (\bu \bu \rho) = - {\bm \nabla} P + {\bm F}\\
    {\partial_t (\rho E)} + {\bm \nabla} \cdot[\bu(\rho E +P)] = 0 \\
    P = (\gamma - 1) \rho U \\
    E = \rho (U + \frac{1}{2} u^2)
  \end{cases}
\end{equation}
where $\rho$ is the mass density, $\bu$ the velocity, $P$ the
pressure, $E$ the total energy density, $U$ the specific internal
energy, and $\gamma$ is the ratio of the specific heats in the
system. Finally, the fourth equation in (\ref{eq:eq}) is the equation
of state for our system. The effect of the large scale forcing ${\bm
  F}$ gives rise to a turbulent flow whose energy is transferred from
scale $L_0$ towards small scales.  The energy input $\int d^3x \, \bu
\cdot {\bm F}$ produces an increase of the internal energy $U$, which
grows in time. One can easily show that the quantity $\int dx^3
P\, \partial_i v_i$ represents the energy transfer
from kinetic to internal energy of the flow.  The sound speed
increases in time as well, while the average Mach number is order
$0.3$. The numerical simulation is done for isotropic and homogeneous
forcing with a resolution $1856^3$. The integration in time have been
done for three  eddy turnover times after a transient evolution. More
details on the numerical code can be found in \cite{prl.chicago}.  The
code develops density
gradients due to compressibility. Despite compressible effects, which
lead to highly non-trivial scaling of density and entropy fluctuations
(\cite{prl.chicago,pouquet}), the net feedback on the velocity scaling
is very weak, if any. Here we show that to the best of our testing
ability, velocity fluctuations in the inertial range are
indistinguishable from those measured on incompressible fluid
turbulence.  \\
\noindent Although the integration is formally inviscid, there is a
net energy transfer from the turbulent kinetic energy $1/2\rho u^2$ to
the internal energy.  Thus we may think that an effective viscosity
$\nu_{\mbox{eff}}$ is acting on the system. In order to estimate it, we
proceed as if the Kolmogorov 4/5 equation (see \cite{Fr.95}), valid
for incompressible turbulence applies to our case --with effective
parameters: \be S_L^{(3)}(r) = -\frac{4}{5} \epsilon_{\mbox{eff}} r + 6
\nu_{\mbox{eff}} \frac{d}{dr} S_L^{(2)}(r).  \ee
A fit of our data with this formula gives,
$\epsilon_{\mbox{eff}}=0.054, \nu_{\mbox{eff}} = 8.3 \cdot 10^{-6}$,
which corresponds to a Kolmogorov scale, $\eta = (\nu_{\mbox{eff}}^3 /
\epsilon_{\mbox{eff}}^{1/4})$, equivalent to roughly half grid cell
and to $R_{\lambda} \sim 600$.
The dynamical effects of the effective viscosity is however different
from what one usually observes in the Navier Stokes equations,
i.e. the dissipation range does not behave similarly to the Navier
Stokes solutions. Thus, while on the energy flux (i.e. the third order
structure functions) we still observe an effective dissipation, the
dissipation range is changed according to the specific mechanism
employed in the simulation, namely the increase of internal energy and
the artificial numerical viscosity added up to smooth too high
gradients (\cite{chicago}).
\begin{figure*}
  \begin{center}
    \includegraphics[scale=0.5]{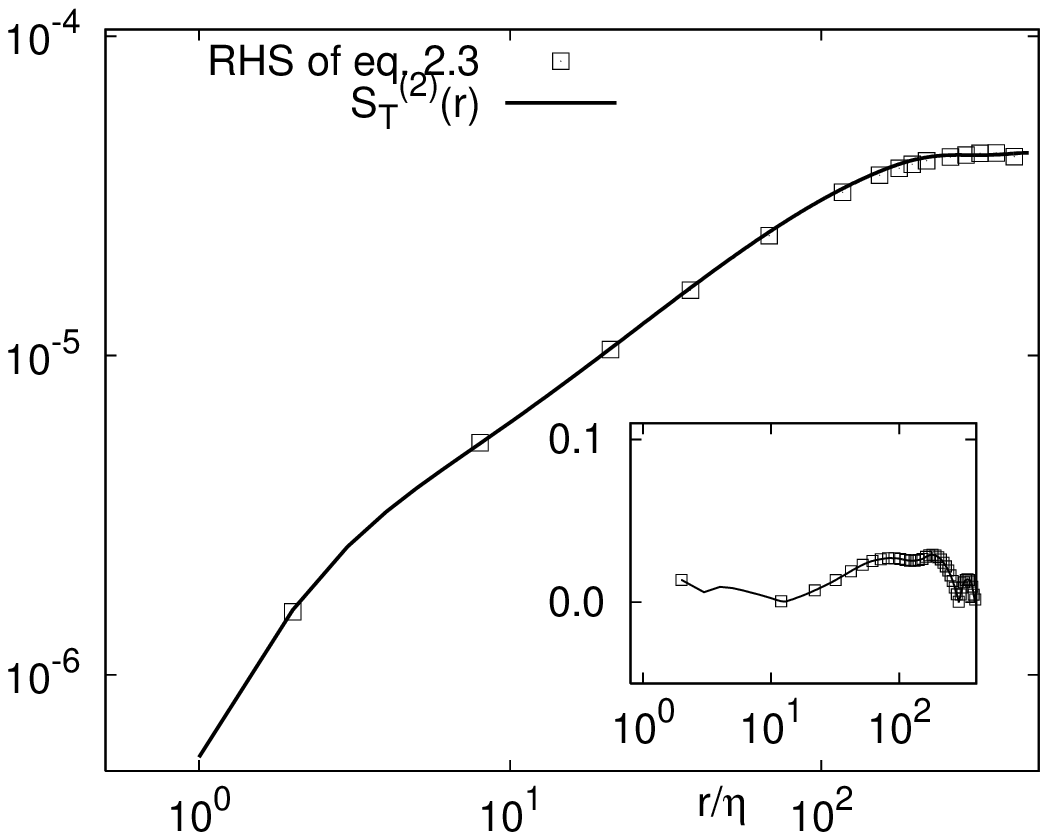}
    \includegraphics[scale=0.5]{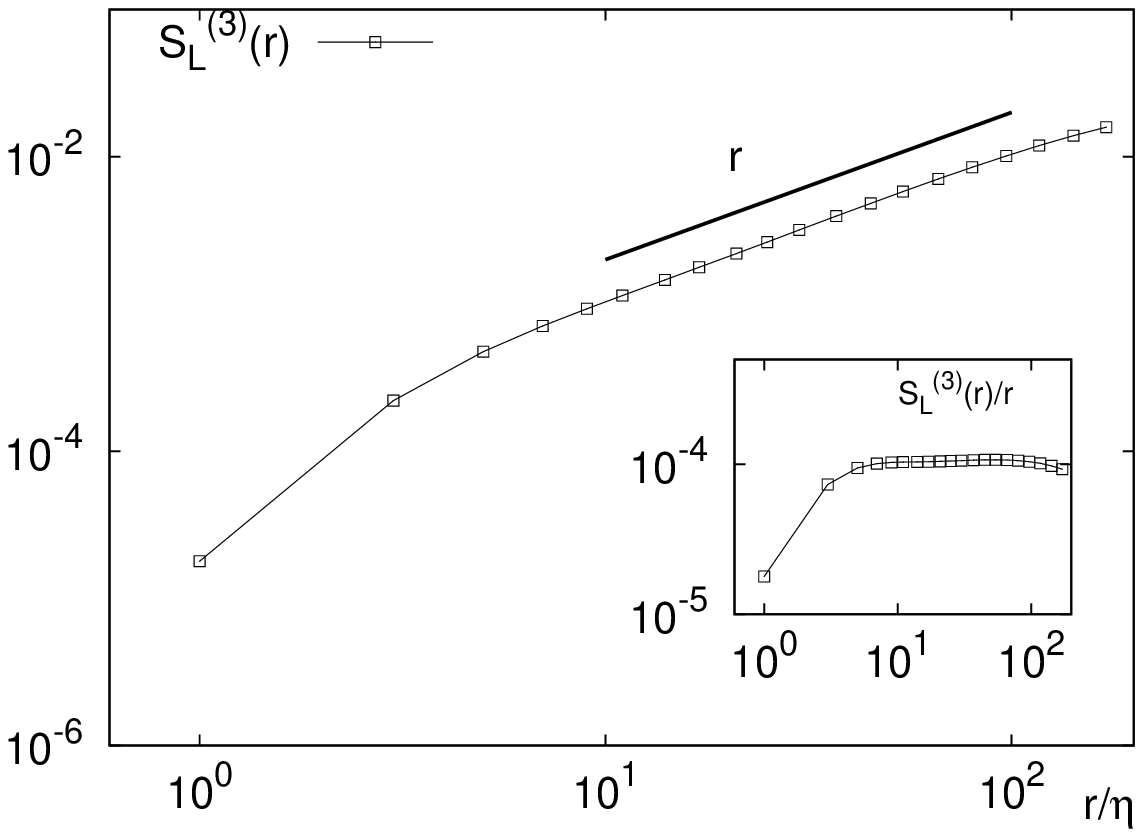}
    \caption{Left: Test of the isotropic and incompressible
      constraint, we show the LHS and the RHS of (\ref{eq:s2ts2l}). In
      the inset we have a percentage estimate of the breaking of the
      relation: (RHS-LHS)/LHS. Notice that percentage-wise the
      relation is well verified, within $5\%$ of
      accuracy. Right: third order longitudinal structure functions,
      $S_L^{(3)}(r)$ with the power law exact relation coming from the
      $4/5$ law (\cite{Fr.95}). Inset: compensated plot,
      $S_L^{(3)}(r)/r$.}
    \label{fig:s2ls2t}
  \end{center}
\end{figure*}
Let us quantify the importance of the small compressibility on the
statistics of the velocity field.  For isotropic and incompressible
fields, there exists an exact relation which connects second order
longitudinal and transverse structure functions (\cite{Fr.95}):
\begin{equation}
  S^{(2)}_T(r) = S^{(2)}_L(r) + \frac{r}{2} \frac{d}{dr} S^{(2)}_L(r),
  \label{eq:s2ts2l}
\end{equation}
this equation is useful because any deviations from it gives a
quantitative hint on the cumulative importance of anisotropy and
compressibility, scale-by-scale.  We show in the left panel of
Fig. (\ref{fig:s2ls2t}) the comparison between $S^{(2)}_T(r)$ and its
reconstruction via the RHS of (\ref{eq:s2ts2l}).  As one can see the
agreement is good. Combined effects due to anisotropy and
compressibility are less
than a few percent for second-order statistics (see also
Fig.\ref{fig:1lb} for higher-order statistics).

\section{Intermittency and anomalous scaling in the Eulerian velocity
  field}
\label{sec:3}
\begin{figure}
   \begin{center}
     \includegraphics[scale=0.5]{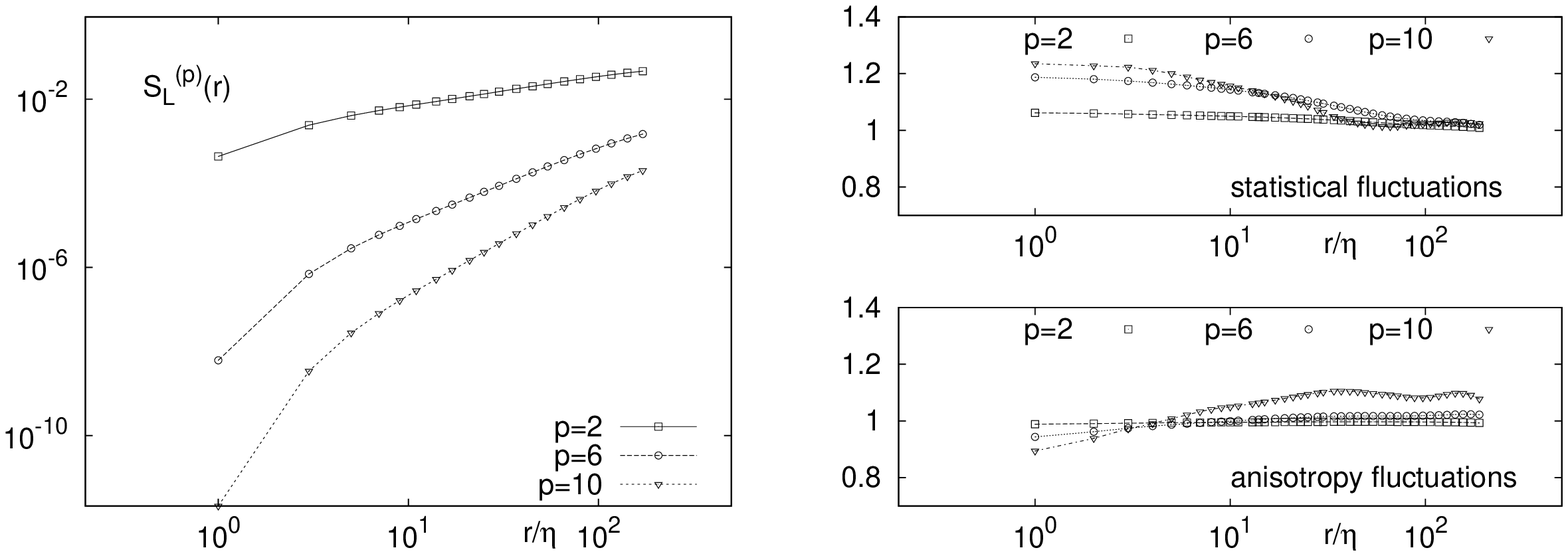}
     \caption{Left: log-log plot of Longitudinal Eulerian Structure
       Functions, $S_L^{(p)}(r)$ for $p=2,6,10$. Right: estimate of
       the anisotropic fluctuations, $\langle
       [u_x(\bx+r_x)-u_x(\bx)]^p \rangle/\langle
       [u_y(\bx+r_y)-u_y(\bx)]^p \rangle$, for $p=2,6,10$ (bottom).
       Estimate of the statistical fluctuations, obtained by plotting
       the ratio of $S_L^{(p)}(r)$ calculated over the whole
       statistics or over one half of it (top). The sum of the two
       error sources leads to error bars that are of the order of the
       symbol size on the left panel.}
     \label{fig:1lb}
   \end{center}
\end{figure}

We start our analysis by measuring the scaling behavior of the
Eulerian structure functions.  In the right panel of
Fig. (\ref{fig:s2ls2t}) we show the third order longitudinal ESF as
measured in our DNS. As one can see, in log-log plot there is a clean
scaling range over more than one decade extending from $ r/\eta \in
[10:100]$, with a slope close to the scaling prediction, $S_L^{(3)}(r)
\sim r$. In Fig. (\ref{fig:1lb}) we show a typical log-log plot of
Longitudinal ESF for $p=2,6,10$. Again, a qualitative scaling
behaviour is clearly detectable. The high resolution and the highly
isotropic forcing used in our DNS, allows also for a precise
quantitative assessment of the importance of statistical and
anisotropy effects on the averaged quantified. In the right panel of
the same figure, we show the estimate of the statistical fluctuations
(top panel) by plotting the ration between two Structure functions
averaged either over the whole statistics or over one half of it. As
one can see, statistical fluctuations become at maximum of order
$10-20\%$ at small scales (where intermittency is more severe) and
only for high order moments $p=6$ and larger. Inside the inertial
range, $r >10 \eta$, they are order $5\%$ or even less. Similarly, in
the bottom panel we give a quantitative estimate of anisotropic
residual effects, by plotting the ratio between longitudinal
measurements over two different orthogonal directions. Here, as
expected, deviations are larger close to the integral scale, reaching
a maximum of $10\%$ for high order moments. As a result, the combined
effect of statistical and anisotropic fluctuations for the Eulerian
scaling are small and they result in an estimate of the error bars
that is within the size of the symbols for the log-log plot showed on
the left panel of Fig. (\ref{fig:1lb}).

\begin{figure}
  \begin{center}
    \includegraphics[scale=0.5]{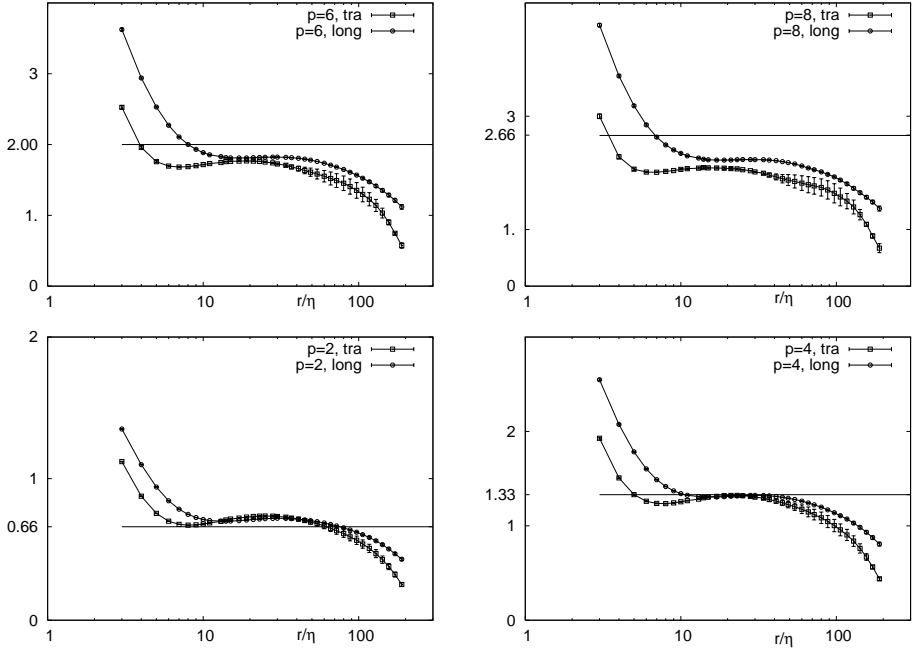}
    \caption{Local scaling exponents, $\zeta_L^{(p)}(r),
      \zeta_T^{(p)}(r)$ of Eulerian Lagrangian and Transverse
      structure functions for $p=2,4,6,8$. Error bars are obtained by
      estimating the residual anisotropy, and they are given by the
      spread between measurements on different directions, $\br/r =
      \hat x, \hat y, \hat z$. Statistical error bars are smaller. The
      horizontal straight line corresponds to the k41 dimensional
      prediction, $\zeta^{(p)}=p/3$.}
    \label{fig:2lb}
  \end{center}
\end{figure}
\begin{figure*}
  \begin{center}
    \includegraphics[scale=0.5]{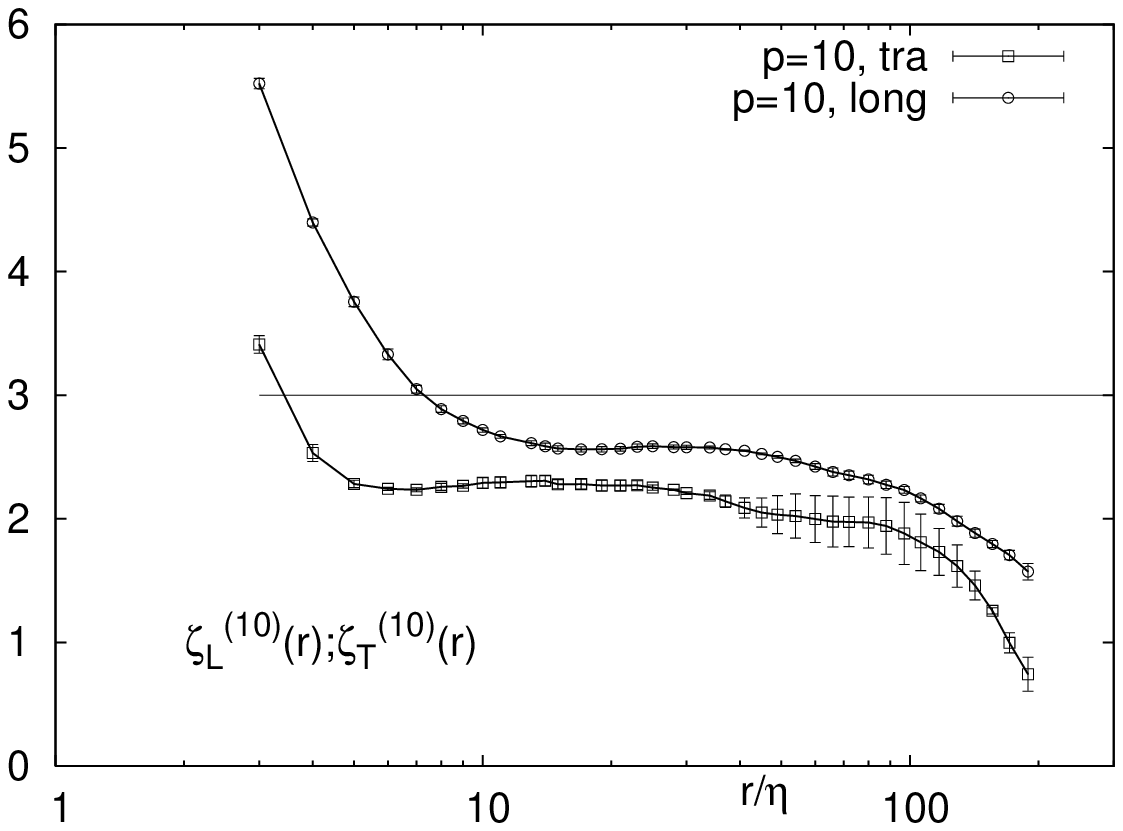}
    \includegraphics[scale=0.5]{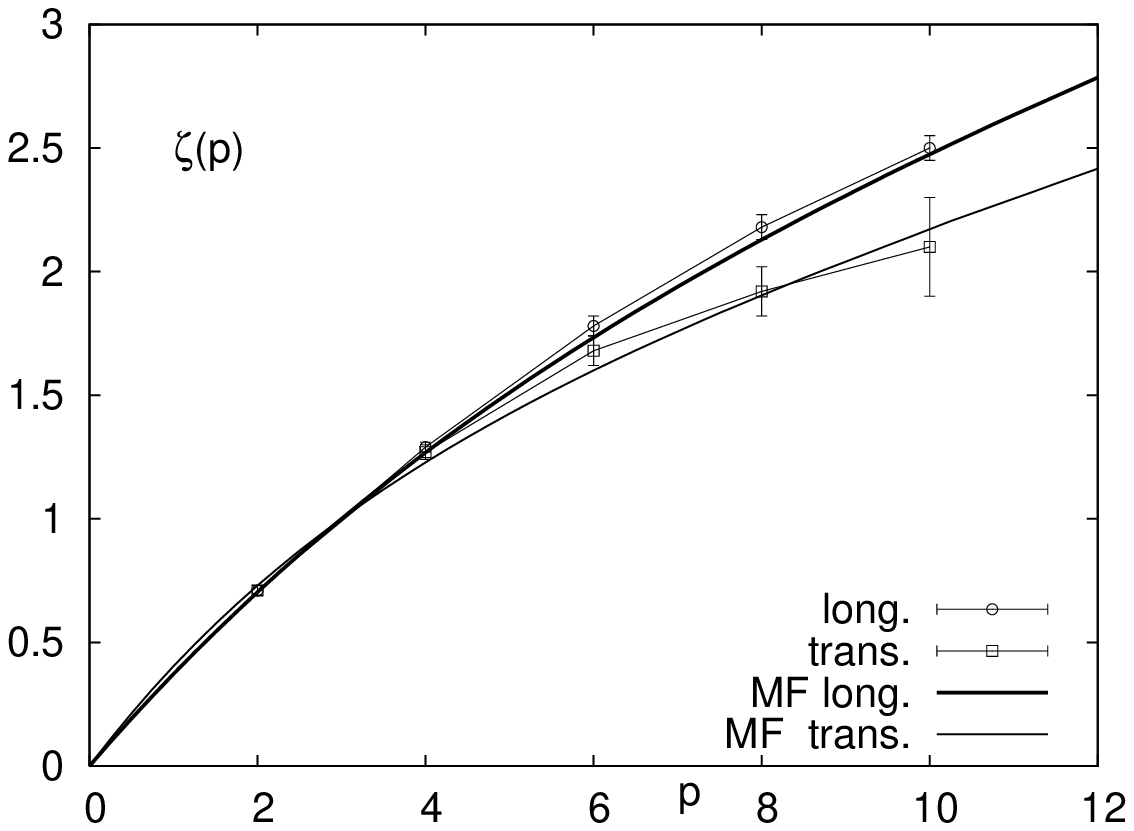}
    \caption{Left: longitudinal and transverse local scaling
      exponents, the same of fig. (\ref{fig:2lb}) but for
      $p=10$. Right: summary of the Eulerian scaling from table 1.
      The two solid lines represent a fit to the scaling exponents
      with two different Multi Fractal spectrum $D_L(h)$ and $D_T(h)$
      both having a log-Poisson statistics, (\cite{sl,dubrulle}):
      $D_{L,T}(h)= \frac{3 (h-h^*)}{ \log(\beta_{L,T})} \left[\log
        \left( \frac{3(h^*-h)}{d_{L,T}^* \log(\beta_{L,T})}\right)
        -1\right] +3 -d_{L,T}^*$.  The free parameters are chosen such
      as to minimize the fitting differences between Longitudinal and
      Transverse. The only parameter we changed is the intermittent
      factor $\beta_L=0.6$ and $\beta_T=0.4$ while we kept the highest
      singularity to be the same for the two cases, $h^*=1/9$. The
      fractal dimension of the $h^*$ is then fixed by the requirement
      that $\zeta_L^{(3)}=\zeta_T^{(3)}=1$ and must be chosen $d^*_{L,T} =
      (1-3h^*)/(1-\beta_{L,T}) $. Another alternative approach for the
      fit would be to fix
      $\beta$ and change the value of the highest singularity between
      longitudinal and transverse fluctuations.  }
    \label{fig:3lb}
  \end{center}
\end{figure*}

One of the pluses of high resolution DNS is the possibility to go
beyond log-log plot, analysing scaling properties locally, with high
precision.  In order to do that, we analyse scaling by using the
magnifying glass of local scaling exponent, (LSE), given by the
log-derivative of Structure Functions: \be \zeta_L^{(p)}(r) \equiv
\frac{d \log S_L^{(p)}(r)}{d \log r}; \qquad \zeta_T^{(P)}(r) \equiv
\frac{d \log S_T^{(p)}(r)}{d \log r}. \ee The advantage to rely on LSE
is twofold. First, it allows for assessing statistical properties
locally, removing large-order non-universal contributions coming from
the overall prefactors in the SF. Second,
 and more importantly, they assess scaling without the need for any
 fitting; in other words, they are just the outcome of a measure and
 they do not depend on an {\it arbitrary} definition as the extension
 of the inertial range, in order to define the degree of
 intermittency. For instance, let us consider, the scaling of fourth
 order longitudinal Flatness. It is easy to see that it can be
 rewritten, independently on any scaling assumption and for all
 separations $r$, as: \be F_L^{(4)}(r) =
 \frac{S_L^{(4)}(r)}{(S_L^{(2)}(r))^2} =
 (S_L^{(2)}(r))^{\phi_L^{(4)}(r)};\qquad \phi_L^{(4)}(r) =
 \frac{\zeta_L^{(4)}(r)}{\zeta_L^{(2)}(r)}-2.\ee In other words, the
 presence of intermittency, i.e. a non trivial scale-dependent
 Flatness behaviour, is directly measure by how much the ratio of LSE,
 $\zeta_L^{(4)}(r)/\zeta_L^{(2)}(r)$ is different from its dimensional
 value, $2$, scale-by-scale.
Let us start to analyze the LSE for the Longitudinal and Transverse
ESF.  In figure (\ref{fig:2lb}) we show the LSE for $p=2,4,6,8$ and in
figure (\ref{fig:3lb}) we show the LSE for $p = 10$.  Let us first
notice that Longitudinal and Transverse statistics have
significantly different viscous cut-off, with transverse fluctuations having
the tendency to have less important viscous damping with respect to
the longitudinal one (\cite{Go.02}). This is an effect already
present for low order $p=2$ (bottom left panel) and therefore
certainly induced, at least partially, by geometrical constraints as
the ones discussed about eqn. (\ref{eq:s2ts2l}). Beside this remark,
let us also notice the small --in amplitude-- but important --in
principle-- difference in the scaling exponents form longitudinal and
transverse. Up to $p=4$ the LSE are almost coinciding, within error
bars, in the range of scales $r/\eta \in [20:80]$. For larger orders,
we start to detect a difference, with the transverse being
systematically below the
longitudinal. This difference becomes significantly visible for $p=10$
as shown separately in Fig. (\ref{fig:3lb}).  Note that this statements
is not the outcome of any fitting procedure and what is shown in
figure (\ref{fig:2lb}) and (\ref{fig:3lb}) is a measurements
independent on any theoretical interpretation.
To decide about the existence --and extension-- of a {\it power-law}
scaling range needs some fitting. The value of the {\it global}
scaling exponent, $\zeta_L^{(p)}, \zeta_T^{(p)}$ would then be given
by the average of the LSE in the region where they are close to
constant. An error bar on the mean value can be then estimated on the
basis of the oscillation induced by statistical and anisotropic
effects in the fitting range and by the change as a function of the
extension of the scaling range used to make the fit. Such a fitting
procedure (details in the caption) leads to the summary for the {\it
  global } Eulerian scaling exponents depicted in the right panel of
Fig. (\ref{fig:3lb}) and summarized also in table 1, where we compare
our results also with other DNS obtained with fully incompressible
Navier-Stokes equations at comparable Reynolds numbers.  From a
theoretical point of view, we know that there are exact constraints
that fix $\zeta_L^{(p)}= \zeta_T^{(p)}$ for $p=2$ (for $p=3$ a similar
constraint fix the scaling of third order longitudinal with mixed
second order transverse and first order longitudinal
averages). Nevertheless, there is not any strong theoretical argument
suggesting the possibility that for $p \ge 4$ one should expect
different scaling. So, the result shown in
Figs. (\ref{fig:2lb}-\ref{fig:3lb}) is not fully understood. Whether
the difference between longitudinal and transverse ESF will shrink
going to higher Reynolds number remains to be investigated, and it is
an important open question.  The good news is that the value found in
our numeric are in perfect agreement with previous results (see table
1), making us confident that they are robust and not dependent on
compressibility. Even the relative scaling behaviour, obtained by
plotting each structure functions versus the third order one, a
procedure known as Extended Self Similarity (ESS) in the literature
(\cite{ess1,ess2}) does not change much the above picture.  In table 1
we give also the results of the LSE for both Longitudinal and
Transverse ESF estimated by using ESS. Again a small spread between
longitudinal and transverse is measured. Nevertheless, it is
interesting to notice that while the ESS procedure applied to
longitudinal statistics reduces considerably the error bars with
respect the usual LSE, for the transverse one there is not a clear
gain in adopting ESS. This may suggest the possibility that the
different scaling behaviour detected between, $S_L^{(p)}(r)$ and
$S_T^{(p)}(r)$ maybe due to different sub-leading corrections to the
main leading scaling, contributing more to the transverse scaling than
to the longitudinal. Due to the lack of any hints on the possible
sub-leading correction we refrain here from entering in to a fitting
procedure with too many free parameters. The issue whether the
different in the LSE shown in Fig. (\ref{fig:2lb}) is the signature of
a true mismatch between the scaling properties of longitudinal and
transverse fluctuations or the result of a superposition of leading
and sub-leading power laws with coinciding exponents between
longitudinal and transverse but with different prefactors, remains
open.

\begin{table*}
  \begin{center}
    \begin{tabular}{c c c c c c c }
      \hline
      \\p & $\zeta_L^{(p)}$ & $ \zeta_T^{(p)}$ & $\zeta_L^{(p)}$ [G02] & $\zeta_T^{(p)}$ [G02] &  $\zeta_L^{(p)}/\zeta_L^{(3)}$ &  $\zeta_T^{(p)}/\zeta_T^{(3)}$ \\ 
      \hline 
      2 &0.71 $\pm$ 0.02 & 0.71 $\pm$ 0.02  & 0.70 $\pm$ 0.01  & 0.71 $\pm$ 0.01 & 0.69 $\pm$ 0.005 & 0.71 $\pm$ 0.01 \\
      4 &1.29 $\pm$ 0.03 & 1.27 $\pm$ 0.05  & 1.29 $\pm$ 0.03 & 1.26 $\pm$ 0.02  & 1.28 $\pm$ 0.01  & 1.26 $\pm $ 0.01\\
      6 &1.78 $\pm$ 0.04 & 1.68 $\pm$ 0.06  & 1.77 $\pm$ 0.04 & 1.67 $\pm$ 0.04 &  1.75 $\pm$ 0.01  & 1.68 $\pm$ 0.03 \\
      8 &2.18 $\pm$ 0.05 & 1.92 $\pm$ 0.10  & 2.17 $\pm$ 0.07 & 1.93 $\pm$ 0.09 &  2.17 $\pm$0.03  & 1.98 $\pm$ 0.1  \\
      10 &2.50 $\pm$ 0.06 & 2.10 $\pm$ 0.20  & 2.53 $\pm$ 0.09  & 2.08 $\pm$ 0.18 &  2.5 $\pm$ 0.05    &  2.25 $\pm$ 0.15

    \end{tabular}
    \caption{Estimate of the global scaling exponents, out of the
      figures for the local scaling exponents
      (\ref{fig:2lb}-\ref{fig:3lb}). The first and second columns
      refer to our DNS, the third and fourth to the data published in
      (\cite{Go.02}) from a DNS at comparable Reynolds number. The
      last two columns correspond to the ESS estimate of our data,
      using the third order longitudinal ESF as a reference for the
      scaling of $S_L^{(p)}$ and using the third order transverse ESF
      for the scaling of $S_T^{(p)}(r)$. Error bars are obtained by
      summing the uncertainty obtained from estimating the scaling in
      three different spatial direction (anisotropy contributions) and
      by changing the scaling range in the interval $ r \in
      [10:100]\eta$, where the global exponent is evaluated.  }
  \end{center}
\end{table*}

\section{Intemittency and anomalous scaling in 
the Lagrangian velocity field}
\label{sec:4}
\begin{figure}
  \centering
  \includegraphics[scale=0.5]{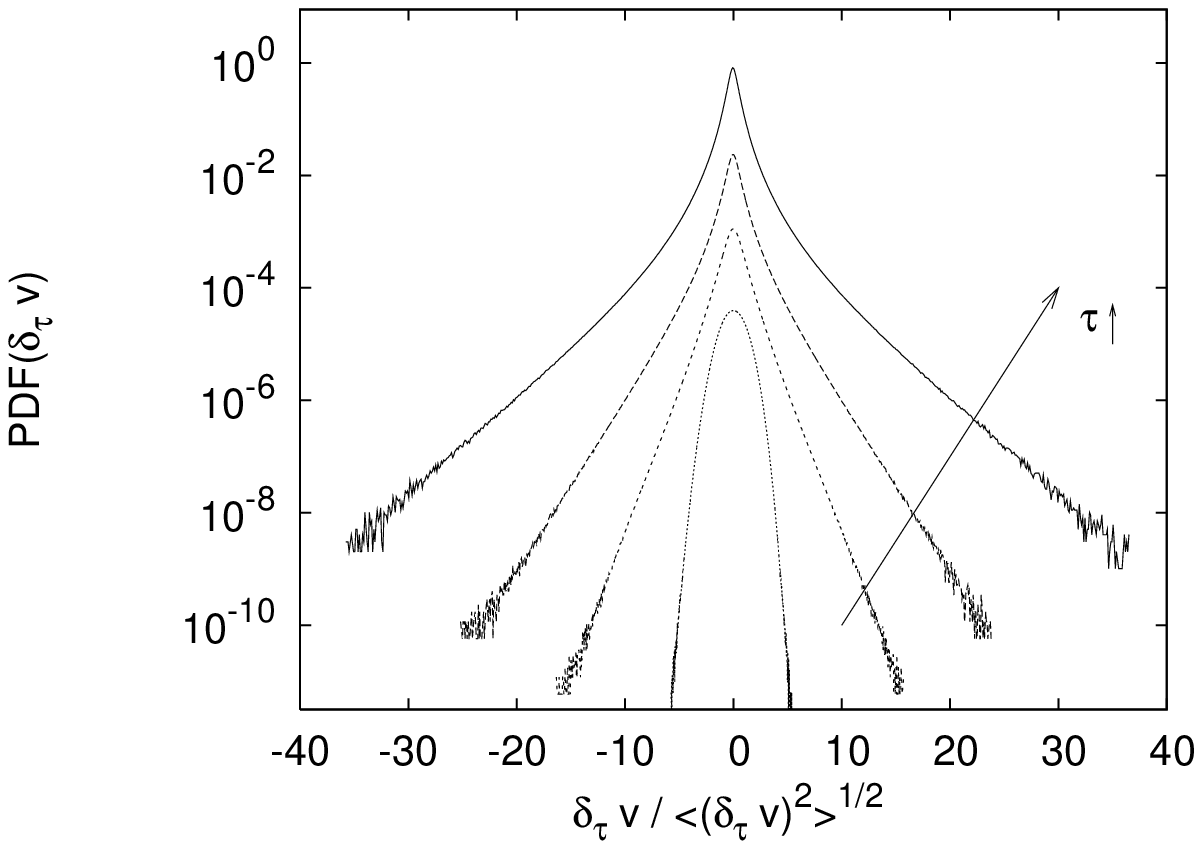}
  \includegraphics[scale=0.5]{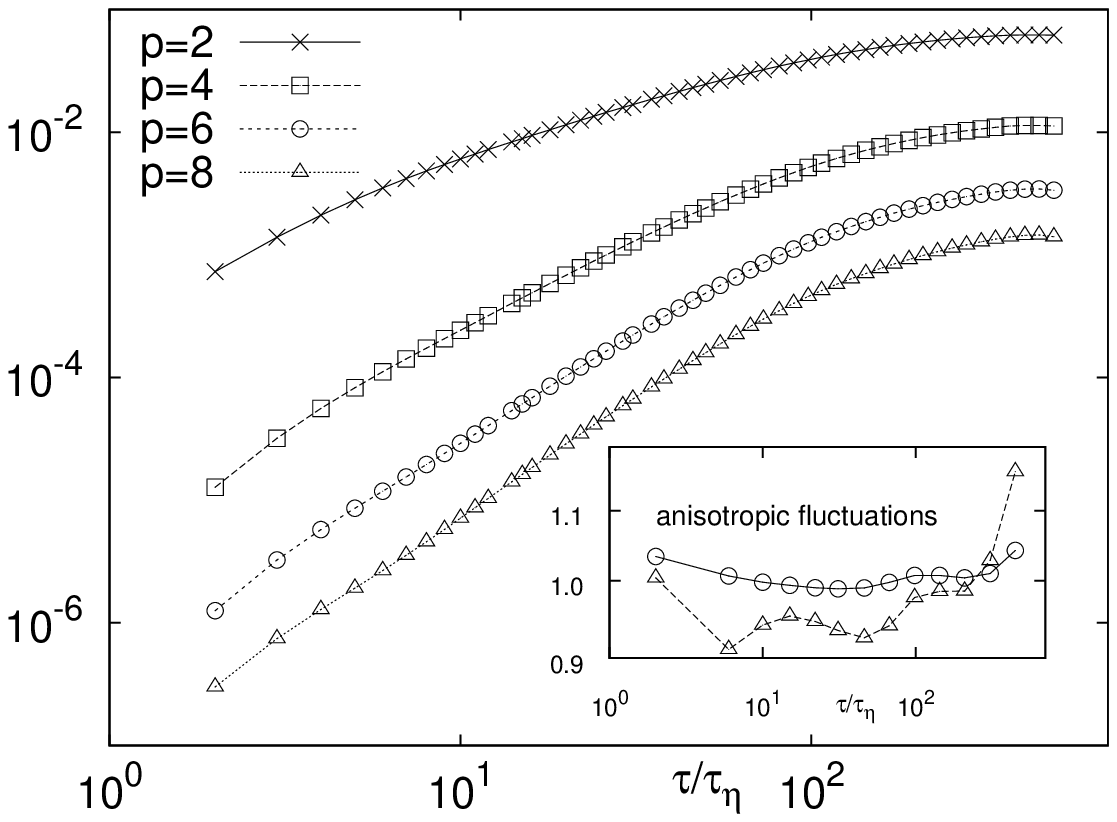}
  \caption{Left: Lagrangian PDF for a single component velocity
    increments along particle trajectories, over different time
    increments, $\tau \in [2:400]\tau_\eta$. Right: log-log plot of
    LSF for $p=2, 4, 6, 8$. Inset: estimate of the anisotropic
    statistical degree: ratio between two different components of LSF,
    $\langle (\delta_\tau v_x)^p \rangle/ \langle (\delta_\tau v_y)^p
    \rangle $, for $p=4$ (circles) and $p=8$ (triangles)}
  \label{fig6}
\end{figure}

We now turn our attention to Lagrangian particles.  As it is known,
LSF shows very strong intermittency.  In figure (\ref{fig6}), we show
the probability density function of $\delta_{\tau} v_i$, for different
values of $\tau$ and averaging over the different velocity components: for small $\tau$ the probability density exhibits
stretched exponential tails, as also measured experimentally in (\cite{MMMP.01}).
The huge statistics at
our hands, up to 16 millions trajectories of length up to one large
scale eddy turn over time, $T_L$, allows for detecting fluctuations up
to $40$ standard deviations, for the highest frequency. Concerning the
LSF, we show in the same figure (right panel) a first overview on a
log-log scale for moments up to $p=8$. From this panel we can already
extract some conclusions.
 First, scaling in Lagrangian framework is not as good as the
Eulerian one, as one can judge by the naked eye.  This is a common
feature of all Lagrangian statistics, and was already observed in many
other experimental and numerical previous works.
The simplest explanation is that the finite Reynolds effects are more
important in Lagrangian domain: dimensional estimate gives for the
Lagrangian inertial range extension the scaling $T_L/\tau_\eta \sim
R_{\lambda}^{1/2}$, while in the Eulerian case we have $L/\eta \sim
R_\lambda^{3/4}$. In the inset of the same figure, we show the
importance of anisotropic fluctuations, by comparing the ration
between two LSF on two different components,
$\cS^{(p)}_x(\tau)/\cS^{(p)}_y(\tau)$ for $p=4,8$. As one can see, the
importance of anisotropic fluctuations is again of the order of $10\%$
at most, for the highest moments. Such correction are of the order of
the symbol size of the LSF averaged over the three components shown in
the left panel. The poorer scaling in the Lagrangian domain, does not
allow for a systematic assessment on the local scaling exponents as
done for the Eulerian field. Here, if one wishes to retain good local
properties, it is necessary to resort to the ESS method, plotting the
local scaling exponents relative to one moment versus a reference one
(here taken the second order): \be \chi^{(p)}(\tau) \equiv \frac{ d
  \log \cS^{(p)}(\tau)}{d \log \cS^{(2)}(\tau)},
\label{eq:chip}
\ee which are related, in presence of pure power law scaling, to the
scaling exponents of the LSF defined from (\ref{eq:scaling}) by the
obvious relation: \be \chi^{(p)}(\tau) \sim const. = \xi^{(p)}/\xi^{(2)}.
\label{eq:22}
\ee Let us stress nevertheless, that the importance of the local
exponents (\ref{eq:chip}) goes much beyond their interpretation as a
simple proxy of the ratio between the structure functions exponents
(\ref{eq:22}). Indeed, they give a clear and simple way to assess the
importance of intermittency in the Lagrangian domain being the only
quantities entering in the scaling properties of Lagrangian
hyper-Flatness: \be \cF^{(p)}(\tau) =
\frac{\cS^{(2p)}(\tau)}{(\cS^{(2)}(\tau))^2} =
(\cS^{(2)}(\tau))^{\chi^{(2p)}(\tau)-p/2}.
\label{eq:fla.lag}
\ee 
Again here the same comment made for the Eulerian case is in order:
via the Lagrangian hyper-Flatness (\ref{eq:fla.lag}) we are able to
assess intermittency in a quantitative way, {\it free of any
fitting ambiguity}, without having to assume
power law properties, by simply checking the difference between
$\chi^{(2p)}(\tau)$ and $p/2$, scale-by-scale.
\begin{figure}
\centering
\includegraphics[scale=0.5]{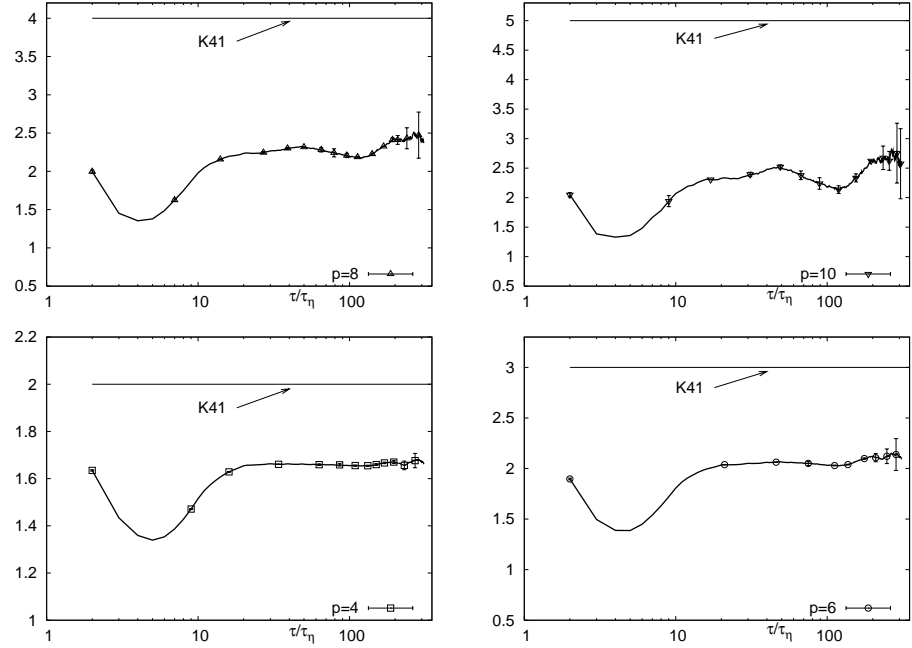}
\caption{Lagrangian Local scaling exponents for hyper-Flatness,
  $\chi^{(p)}(\tau)$, for $p=4,6,8,10$. Notice the extremely good
  scaling behaviour for low moments, which deteriorates only mildly
  for $p=6,8$.  Notice also the dip region close to the viscous time
  scales, an effect interpreted in terms of trapping into vortex
  filaments (\cite{jmenez,BBCLT.06}). The horizontal straight lines correspond
  to the dimensional scaling $\chi^{(p)}= p/2$.}
\label{fig:lse}
\end{figure}
\begin{table*}
  \begin{center}
    \begin{tabular}{c c c c c}
      \hline
      \\    $\chi^{(p)}$  &  $p=4$  & $p=6$  &  $p=8$  & $p=10$\\ 
      \hline
      our DNS ($R_\lambda \sim 600$)  &1.66 $\pm$ 0.02 & 2.10 $\pm$ 0.10  & 2.33 $\pm$ 0.17 & 2.45 $\pm$ 0.35 \\
      EXP1 ($R_\lambda \sim 800$)    &1.47 $\pm$ 0.18 & 1.73 $\pm$ 0.25  & 1.92 $\pm$ 0.32 & 1.98 $\pm$ 0.38 \\
      EXP2 ($R_\lambda \in [500:1000]$) &1.56 $\pm$ 0.06 & 1.80  $\pm$ 0.20  & --            & -- \\
      DNS1 ($R_\lambda \sim 320$) &1.51 $\pm$ 0.04 & 1.76 $\pm$ 0.11  & --             & --  \\
      DNS2 ($R_\lambda \sim 75$) &1.56 $\pm$ 0.03  & 1.82 $\pm$ 0.08 & 1.92 $\pm$ 0.14 & 1.93 $\pm$ 0.3  \\
    \end{tabular}
    \caption{First raw: summary of the local scaling exponents for
      Lagrangian statistics as extracted from our data shown in
      Fig. (\ref{fig:lse}). Error bars are evaluated out of the
      anisotropic statistical degree, estimating the LSE out of the
      three different velocity components and by changing the fit
      range in the interval $\tau \in [20:200]$. The other rows correspond to other data available for high order moments  
from both  experiments at comparable Reynolds numbers and numerical works
      at smaller Reynolds numbers. EXP1 (\cite{X06});  EXP2 (\cite{mordant}); DNS1 (\cite{mueller}); 
      DNS2 (\cite{mordant}). Notice how some of the previous
      data slightly underestimate the value of the scaling
      exponents for $p=6,8,10$. This is due to the fact that,
      due to the limitations either in the statistics or in the Reynolds number, it was not possible to
      assess scaling properties scale-by-scale as done here. For
      example, the fit leading to the numbers in \cite{X06} and in
      \cite{mueller} (second and fourth raw) was done in the range
      $\tau \in [3:6] \tau_\eta$, well inside the dip region. Similarly, the Reyolds number of the numerical data in
 (\cite{mordant}) (fifth row) is very low, probably the strong saturation in the values for $p=6,8,10$ reported there is due again
to the presence of the dip region spoiling the scaling in the short inertial range present at $R_\lambda \sim 75$.} 
  \end{center}
\end{table*}
In Fig. (\ref{fig:lse}) we show the hyper-Flatness local exponents for
$p=4, 6, 8, 10$. The scaling range now is good, for
$\tau/\tau_\eta \in [20:200]$. As one can see, there are two
remarkable facts. First, Lagrangian scaling is much more intermittent
then the Eulerian one, if measured as a deviations from the
dimensional scaling, $\chi(p)=p/2$. For example, for $p=10$, we
measure $\chi^{(10)} = 2.45 \pm 0.35$ which is $50 \%$ off from the
dimensional value $\chi^{(10)}=5$. The second fact is the strong dip
measured across the viscous scale, for time lags
$\tau/\tau_\eta \in [1:10]$. This is similar to the Eulerial 
``bottleneck'' (\cite{lohse}), while for
Lagrangian structure functions was observed first in (\cite{LM.04})
and interpreted as induced by the presence of small scale vortex
filaments in \cite{BBCLT.05,BBCLT.06}. The value of the Lagrangian
exponents measured in our DNS ar reported in table (2) and compared
with previous measurements on other DNS or experiments. Let us stress
that there are not other measurements with comparable statistics and Reynolds number,as
the one here presented for high order moments.  A global fit in log-log
of the scaling properties does not allow to disentangle the
inertial range from viscous-inertial contribution (affected by the
dip) as they emerge from fig. (\ref{fig:lse}).  This explain
the slightly systematic underestimate of the scaling exponents from
all the previous studies with respect to our measurements. The only
data presented previously for $p=8,10$ at comparable Reynolds  are those published in
(\cite{X06}) where the scaling properties where estimated very close,
if not inside, the dip region, $\tau \in[3:6]\tau_\eta$. Similarly was
done for the data in \cite{mueller}. The numerical data for $p=8,10$ presented in (\cite{mordant}) 
are also understimating the value of the scaling properties with respect to  our data. 
Probably, in this case, the very low value of the  Reynolds number ($R_\lambda \sim 75$)  
is such that the whole scaling region is dominated by the viscous dip.
Considering all these pitfalls,
we may conclude that when scaling properties are analysed in the
correct range and where statistical fluctuations are not too high,
there exists a good universality in Lagrangian statistics, as also
tested quantitatively for $p=4,6$ in
\cite{biferaletal07,arneodoetal08}. 
As for  universality of high order moments, from $p=8$ and up, we need to wait for 
other data with statistical properties and Reynolds number comparable with those of this study, 
to decide.

\section{Multifractal: a link between Eulerian and Lagrangian
  statistics}
\label{sec:5}
Multifractals have been introduce 25 years ago to explain deviations
to the K41 scaling for Eulerian isotropic turbulent fluctuations
(\cite{ParisiFrisch}, see also \cite{review.angelo} for a recent
review).  Since those works, they have been extended to describe also
velocity gradients (\cite{Ne.90,Bif.91}); multi-scale velocity
correlations (\cite{proc,bif_fs}) and  fluctuations of the Kolmogorov
viscous scale (\cite{PV.87,meneveau,lb1,FV.91}).  Let us notice that other attempts have been made to introduce fluctuations of the Kolmogorov scale in turbulence (\cite{lvov,yakhot1,yakhot2}), most of them are equivalent or give anyhow  results almost  undistinguishable from the multifractal approach followed here (\cite{benzi.bif}).

In principle, one is free to
develop different, uncorrelated, multifractal description for Eulerian
and Lagrangian quantities (\cite{CRLMPA.03}). On the other hand, in
reality, Eulerian and Lagrangian measurements are of course intimately
linked. One is therefore naturally tempted to develop a {\it unified}
Multifractal description, able to describe both temporal and spatial
fluctuations (\cite{Bo.93,bof02}). In the following we describe such
development starting from the inertial range and finally for the
viscous range. 
\subsection{Inertial range}
Concerning Eulerian MF in the inertial range, the idea goes back to
\cite{ParisiFrisch} and reads as follows. Suppose we have different
velocity fluctuations with different local Holder exponents, $\delta_r
u \sim r^h$, for scale separation in the inertial range. Suppose that
the set where the velocity field has a $h$-exponent is a fractal set
with dimension $D(h)$. Then the ESF can be easily rewritten as an
ensemble average over all possible $h$ fluctuations: \be
\label{eq:emf}
\langle (\delta_r u )^p\rangle = \int_h dh r^{hp} r^{3-D(h)} \sim
r^{\zeta^{(p)}} \ee where the factor $r^{3-D(h)}$ gives the
probability to fall on the fractal set with $h$-exponent at scale $r$
and where we have neglected, for the moment, any difference between
longitudinal and transverse, calling the generic Eulerian increment,
$\delta_r u$.  The last passage in (\ref{eq:emf}) is obtained in the
saddle point limit $r \rightarrow 0$: \be \zeta^{(p)} = \min_h\left(hp
  +3 -D(h)\right).
\label{eq:sp} \ee It is easy to imagine many different possible $D(h)$
distribution --built in terms of random cascade-- leading to a set of
exponents $\zeta^{(p)}$ close to those measured. In previous section
we have shown that both our DNS and previous DNS suggest the
possibility to have different scaling exponents for Longitudinal or
Transverse SF. Therefore we must allow to have two different set of
fractal dimensions, $D_L(h)$ and $D_T(h)$ describing the statistics.
For example, in the right panel of Fig. (\ref{fig:3lb}) we show the
result of the saddle point (\ref{eq:sp}) obtained with two different
fractal spectra, tuned to fit the empirical scaling
exponents as found from our DNS (see caption for details).\\
One possible simple way to link the Eulerian and Lagrangian statistics
is to follows the ideas in \cite{Bo.93}, which we summarize
hereafter. As far as scaling is concerned, in 3d turbulence we may
imagine that the only relevant time in the inertial range is the local
eddy turn over time: \be \tau(r) \sim r/(\delta_r u) \sim r^{1-h}.
\label{eq:tra}
\ee 
Then, we may assume that the Lagrangian velocity increment over a
time lag $\tau$ must be of the order of the corresponding Eulerian
velocity increments over a scale $r$ with $r$ and $\tau$ connected by
(\ref{eq:tra}): \be \delta_\tau v \sim \delta_r u\qquad \tau =
r^{1-h}.  \ee By using only this simple statement one is able to link
now the LSF to the ESF, by simply writing: \be \langle (\delta_\tau v
)^p\rangle = \int_h dh \tau^{\frac{h p}{1-h}} r^{\frac{3-D(h)}{1-h}}
\sim \tau^{\xi^{(p)}}, \ee with now the Lagrangian exponents given by
the saddle point estimate: \be
\label{eq:llse}
\xi^{(p)} = \min_h\left(\frac{hp +3 -D(h)}{1-h}\right).  \ee It is
important to notice that via the translation factor (\ref{eq:tra}) we
obtain two different sets of Lagrangian and Eulerian exponents but
given by the same $D(h)$ curve. The exponents are different because
the local eddy-turn-over is itself a fluctuation quantity, depending on the
local scaling exponent, $h$.

\begin{figure}
  \begin{center}
    \includegraphics{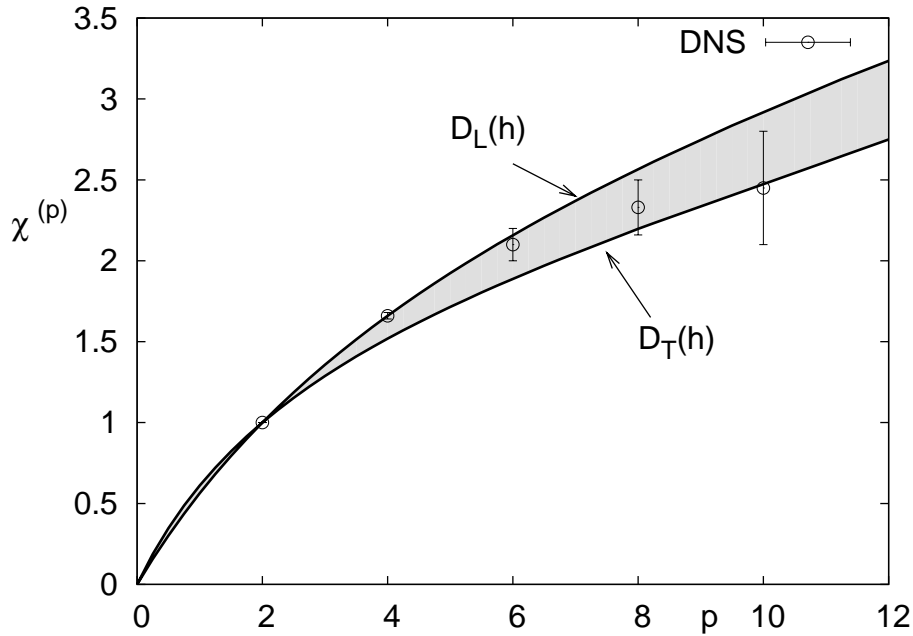}
    \caption{Summary of the Lagrangian Scaling exponents as measured
      in our DNS in the inertial range of time scales (circles),
      together with the prediction obtained from the Eulerian
      statistics by using the bridge relation either with the
      longitudinal Eulerian statistics (upper limit of the shadowed
      area) or with the transverse one (lower limit of the shadowed
      area). }
    \label{fig.mf0}
  \end{center}
\end{figure}

Once given the $D(h)$,extracted from a  fitting of the Eulerian statistics, it is tempting
to use it as an input in (\ref{eq:llse}) to get a prediction with {\it
  no free} parameters for the Lagrangian scaling. Indeed, because the
Eulerian statistics is different depending if one takes longitudinal
or transverse fluctuations, and because it is very natural to think
that during the Lagrangian evolution both fluctuations are felt, one
may imagine to get two different prediction for the Lagrangian
exponents: \be \xi^{(p)} = \min_h\left(\frac{hp +3
    -D_L(h)}{1-h}\right);\qquad \xi^{(p)} = \min_h\left(\frac{hp +3
    -D_T(h)}{1-h}\right).
\label{eq:pre}
\ee depending which Eulerian statistics one uses, $D_L(h)$ or
$D_T(h)$. In fig.~(\ref{fig.mf0}) we show the comparison between the
empirical Lagrangian exponents as extracted from our DNS (see table 2)
and Fig. (\ref{fig:lse}) and the two prediction one get using the
first or second expression in (\ref{eq:pre}).  As one can see the
agreement is  good, indicating that the simple, but non trivial,
bridge relation (\ref{eq:tra}) is working well and capturing the main
scaling properties of both Eulerian and Lagrangian statistics. From
figure (\ref{fig.mf0}) one may see that only for the highest order
$p=10$ one start to see a trend deviating from what predicted by the
Eulerian-Lagrangian bridge relation, although within error bars it
still holds. If this is the signature that very intense Lagrangian
fluctuations bring new information beyond the ones collected by
(\ref{eq:tra}) is an open and interesting question, which is not
possible to address till new empirical and numerical data will allow to access
higher order statistics for both Eulerian and Lagrangian domains.

\subsection{Dissipative Effects}
\begin{figure}
  \begin{center}
    \includegraphics[width=1.\hsize]{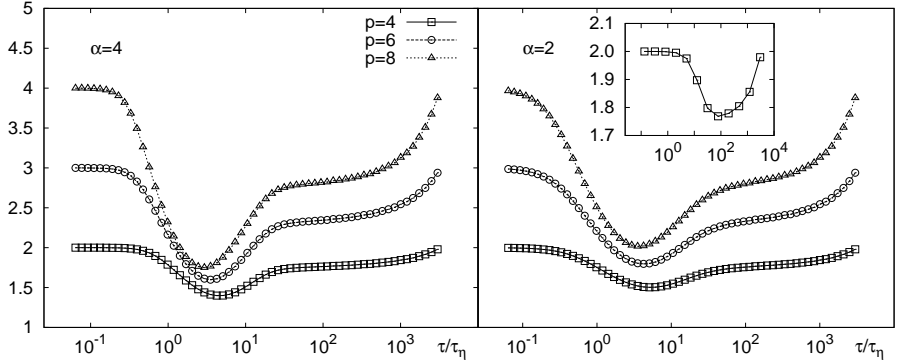}
  \caption{Results for the local Lagrangian exponents,
    $\chi^{(p)}(\tau)$ as obtained from the finite Reynolds
    multifractal formula (\ref{eq:fitL}-\ref{eq:pL}-\ref{eq:lsfmulti})
    for a given Reynolds number and $p=4,6,8$, at changing the free
    parameter, $\alpha$, that sets the importance of the dip
    region. Left: $\alpha=4$; Right: $\alpha=2$. In the inset of the
    right panel we also show, $\chi^{(4)}(\tau)$ obtained from the
    same multifractal formula but keeping fixed the dissipative time
    scales, $\tau_\eta(h) = const.$, in this latter case the dip
    region disappears.  }
  \label{fig.mf1}
 \end{center}
\end{figure}

The possibility to assess Lagrangian data, both numerically and
experimentally, opened the way to investigate also dissipative and
sub-dissipative time scales.  For Eulerian statistics is very
difficult to have reliable data at scales smaller then $\eta$, due to
either limitations induced by the probe size, for experiments, or by
the numerical resolution (see  \cite{joerg2,yakhot.sreene,watanabe,yamazaki} for
recent DNS meant to address this issue). For
Lagrangian quantities, the development of non-intrusive experimental
techniques (mainly based on fast CCD \cite{toschibod}) and the
advancements of numerical algorithm to track particles, allowed for
assessing time fluctuations of the order of $0.1 \tau_\eta$ even at
high Reynolds numbers.  Now, new question about the statistical
fluctuations for $\tau \sim \tau_\eta$ or smaller, arise. Recently, a
lot of attention has been payed on the dip region visible in the local
Lagrangian scaling exponents as shown in fig. (\ref{fig:lse}) both
from numerical and experimental studies (\cite{arneodoetal08}). In the last twenty years, a
lot of work has been done to include viscous fluctuations in
the multifractal theory -- initially for Eulerian statistics--  
\cite{PV.87,Ne.90,meneveau,FV.91,Bif.91}, and
more recently also for Lagrangian statistics (\cite{CRLMPA.03,BBCDLT.04}).  
A simple way to accomplish this is to take into account of the fact
that, according to the MF theory, the viscous scale is not a fixed
homogeneous quantity, but fluctuates wildly due to
intermittency. Moreover, viscous and inertial fluctuations are linked
by the requirement that viscous effects happens when the local
Reynolds number becomes ${\cal O}(1)$ as first remarked by
\cite{PV.87}: 
\be
\label{pv}
\frac{\delta_\eta u\; \eta}{\nu} \sim {\cal O}(1); \qquad \eta^{1+h}
\sim \nu; \qquad \eta(h) \sim Re^{-1/(1+h)}, \ee where we have taken
for simplicity, $L=1,U_0=1$ and $\nu = Re^{-1}$.  Using the bridge
relation (\ref{eq:tra}) we may immediately find out also the
equivalent fluctuations relation for the Kolmogorov dissipative time (\cite{BBCDLT.04}):
\be \tau_\eta(h) \sim \eta^{1-h} \sim Re^{(h-1)/(1+h)}. \ee Taking in
to account this fact, one may use a Batchelor-Meneveau like
parametrisation (\cite{meneveau,yakhot_old,CRLMPA.03,arneodoetal08}) of velocity
increments for all time lags, including the fluctuating viscous time
in the game:
\begin{equation}
\label{eq:fitL}
\delta_{\tau} v  = V_0
\,\frac{\tau}{T_L}\left[\left(\frac{\tau}{T_L}\right)^{\alpha} +
  \left(\frac{\tau_{\eta}(h)}{T_L}\right)^{\alpha}\right]^{\frac{2h-1}{\alpha(1-h)}}\,,
\end{equation}
where $\alpha$ being a free parameter controlling the crossover around
$\tau\!  \sim\! \tau_{\eta}$, and $V_0$ the root mean square large
scale velocity.  It is easy to realize that the above expression has
the correct asymptotic behaviour for very small time lags $\delta_\tau
v \sim \tau a$, where $a = \delta_{\tau_\eta} v \tau_\eta$ is the
acceleration; and in the inertial range, $\delta_\tau v \sim
\tau^{h/(1-h)}$, as described in details in \cite{arneodoetal08}.  In
order to get a prediction for the behaviour of the LSF we need also to
introduce dissipative effects in the MF probability. This can be done
as follows:
\begin{equation}
  \label{eq:pL}
  P_h(\tau,\tau_{\eta}) = {\cal Z}^{-1}(\tau) \left[
    \left(\frac{\tau}{T_L}\right)^{\alpha} +
    \left(\frac{\tau_{\eta}(h)}{T_L}\right)^{\alpha}\right]^{\frac{3-D(h)}{\alpha(1-h)}},
\end{equation}
where ${\cal Z}$ is a normalizing function and $D(h)$ the fractal
dimension of the support of the exponents $h$ (either $D_L(h)$ or
$D_T(h)$ if one wants to link the Lagrangian dissipative physics to
the Eulerian one). At this point, given the Reynolds number, we are
left with two parameters ($\alpha$ and a multiplicative constant in
the definition of $\tau_\eta$).  In order to compute the LSF at all
time lags one needs to integrate the expression (\ref{eq:fitL}) with
the weight given by (\ref{eq:pL}) over all possible $h$-values:
\begin{equation}
  \label{eq:lsfmulti}
  \langle (\delta_{\tau} v)^p \rangle \sim  \int_{h_{min}}^{h_{max}} dh P_h(\tau) [\delta_{\tau} v(h)]^p.
\end{equation}
Clearly, the above picture goes back to the inertial values when we
have $\tau/T_L \rightarrow 0$ and $\tau \gg \tau_\eta$, i.e. when we
can neglect viscous effects and we can perform the saddle point
estimate. Here we do not want to do that, keeping all corrections, and
compare the simple parametrisation (\ref{eq:fitL}), dressed with the
MF theory, with our data.  First, let us see what are the qualitative
results obtained from (\ref{eq:fitL}) at changing the free parameters
on it. In fig. (\ref{fig.mf1}) we show the results for
$\chi^{(p)}(\tau)$ as obtained from the global MF description
(\ref{eq:lsfmulti}) for $p=4,6,8$ with a fixed $D(h)$ at changing the
free parameter $\alpha=4$ (left panel) or $\alpha=2$ (right panel).
From this figure we learn two things. The MF description is
very close, already qualitatively speaking, to the real data --compare
with panel corresponding to $p=4$ of Fig. (\ref{fig:lse}). In
particular, the dip present in the data is also captured by the
MF. Moreover, at changing the free parameter $\alpha$ one may deplete
or enhance the dip intensity, as can be see comparing left and right
panel of Fig. (\ref{fig.mf1}).  Let us also notice that the dip region
is entirely due to the presence of the fluctuating viscous cut-off,
$\tau_\eta(h)$, as can be seen in the inset of right panel of
Fig. (\ref{fig.mf1}), where the same data but with a fixed $\tau_\eta$
are plotted and the dip has disappeared.

\begin{figure*}
  \begin{center}
    \includegraphics[scale=0.5]{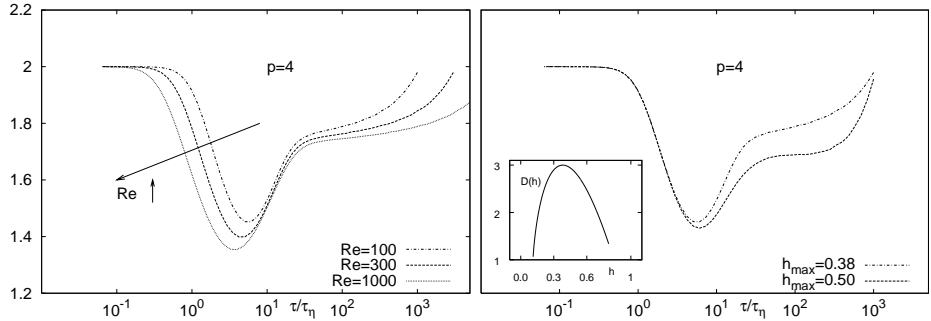}
    \caption{Left panel: Reynolds effects on the local exponents
      predicted from the MF interpolation formula
      (\ref{eq:fitL}-\ref{eq:pL}-\ref{eq:lsfmulti}), we kept fixed
      everything except the Reynolds numbers. Right: sensitivity to
      the whole shape of the $D(h)$ curve of the $\chi^{(4)}(\tau)$ as
      predicted from the MF interpolation formula. We show two
      cases. One case is obtained integrating only on fluctuations in
      the left region of the $D(h)$ curve with respect to its peack, $h
      \in [h_{min},h_{max}]$, with $h_{max}=h_{peack}=0.38$, the second
      case consider also smoother fluctuations up to $h_{max}=0.5$.The
      typical log-Poisson shape of the $D(h)$ used is depicted in the
      inset.}
    \label{fig.mf2}
  \end{center}
\end{figure*}

Second, from fig. (\ref{fig.mf1}) one sees that for any finite Reynolds
numbers, there are corrections to scaling induced by the large time
lags $T_L$; as a consequence, in the inertial range the local
exponents are not exactly constant. In order to give a qualitative
assessment of the importance of this finite Reynolds corrections as
described by formula (\ref{eq:fitL}) we plot in the left panel of
fig. (\ref{fig.mf2}) the values of $\chi^{(4)}(\tau)$ for three
different Reynolds numbers. As one can see, at increasing Reynolds,
the inertial range values becomes flatter and flatter, as they must
do. At difference from what happens in the infinite Reynolds number
limits, for any finite Reynolds, the whole support of $D(h)$ will play
a role in the integration (\ref{eq:lsfmulti}): for not too small
$\tau/T_L$ the saddle point estimate (\ref{eq:llse}) becomes less and
less accurate. As a result, the whole shape of
the $D(h)$ curve becomes important for any order. \\
Indeed, the typical $D(h)$ curve from any MF theory is a convex
function like the one used to fit the Eulerian Longitudinal exponents
depicted in the inset of right  panel of fig. (\ref{fig.mf2}).  The
left part of $D(h)$ curve, with $ h \in [h_{min},h_{peack}]$, is
connected to the positive order of the Eulerian Structure Functions
(\cite{Fr.95}). The right side, $h \in [h_{peack},h_{max}]$, is
connected to negative moments. Negative moments are generally hill
defined and one needs to use inverse statistics to assess the right
part of the $D(h)$ curve (\cite{jensen,bif.inv}). Because of that, we
have a  good control only the left side, where most of the
previous experimental and numerical studies focused. The extension of
the integration in (\ref{eq:lsfmulti}) over the $h$-value on the right
of the peack of $D(h)$ is therefore not based on solid experimental or
numerical basis. The extension to the right part of a given functional
form for $D(h)$ (log-Poisson, log-Normal etc. ), obtained by fitting
only positive moments of Structure functions, is arbitrary.

In the right panel we show the dependence of the final expression
(\ref{eq:lsfmulti}) from the extension of the integration over the $h$
range. As one can see, the inclusion of $h$-value falling on the right
of the peack gives an important contributions to the behaviour of
the local slopes for $\tau \sim T_L$, confirming the importance of the
whole $D(h)$ if one want to control the finite-Reynolds corrections to
the saddle point estimate (\ref{eq:llse}). 
 For a
more detailed investigation of the property of the MF formulation
(\ref{eq:lsfmulti}) also compared with other DNS and experimental
results, the reader can consult the recent collection of data
published in (\cite{arneodoetal08}).

\section{Conclusions}
\label{sec:6}
We have reported about Eulerian and Lagrangian statistics from high
resolution Direct Numerical Simulations of isotropic weakly
compressible turbulence.  Reynolds number at the Taylor microscale is
estimated to be around $600$.  The high number of particle
trajectories explored, together with the high resolution Eulerian flow
allowed us for the first time to study Lagrangian intermittency up to
moments of order $p=10$, at large Reynolds numbers.  The four  main results we have presented
are: (i) Eulerian Longitudinal and Transverse statistics have slightly
different scaling properties, in agreement with previous
incompressible DNS data at comparable resolution. (ii) Lagrangian
statistics is more intermittent than Eulerian ones, percentage-wise
with respect to the observed deviations from dimensional scaling and
its quantitative value can be captured by using a simple multifractal
bridge relations. (iii) In order to correctly asses Lagrangian scaling properties is mandatory to make a scale-by-scale analysis, in order to disentangle the dip region close to viscous scales from the inerital range interval. A lack of such analysis may lead to an underestimate of the Lagrangian scaling exponents, because of the strong intermittency of the dip interval, dominated by viscous effects.(iv) Multifractal phenomenology, is  able to
describe in a unified way:  dissipative effects, inertial range
fluctuations  and large scale corrections (\cite{arneodoetal08}).  \\
This study leaves open a few important questions. First, we need to
wait for higher
Reynolds numbers data, in highly isotropic ensemble, to resolve the
riddle about Longitudinal versus Transverse Eulerian
fluctuations. Second, we need to collect also data with higher
Lagrangian statistics to understand whether the Eulerian-Lagrangian
bridge relation remains valid for higher moments. Third, it is
important to improve experimental and numerical accuracy to measure
small-scales and small-time fluctuations, where highly non trivial
physics is developing as shown by the strong enhancement of local
intermittency in the dip region $\tau \in [1:10]\tau_\eta$. Such
strong enhancement of fluctuations around the viscous scale is due to
local fluctuations of the dissipative cut-off, which reflects in to
the existence of different viscous effects for different moments and
different correlation functions (\cite{FV.91,lvov,yakhot1,benzi.bif}). It is
well described, within the multifractal theory, by the
Paladin-Vulpiani relation (\ref{pv}), as shown by the fact that the
dip region disappear by removing the viscous fluctuations in
(\ref{eq:fitL}-\ref{eq:pL}), see inset of the right side of
fig. (\ref{fig.mf1}). Such effect is probably the highest difficulty
to overcome in both stochastic modeling (\cite{saw2001}) and
 theory of Lagrangian turbulence (\cite{yakhot,zymin}).

We acknowledge very useful discussions with M. Cencini, A. Lanotte,
L. Kadanoff and V. Yakhot. L.B., R.B. and F.T.  thank
 the Flash Center at the University
of Chicago for hospitality when part of this work was prepared. We
also want to thank B. Gallagher for helping in the data analysis and
data rendering.

\end{document}